\documentclass[aps,pra,preprint,floatfix, superscriptaddress]{revtex4-1}

\usepackage{xcolor}
\usepackage{appendix}
\usepackage{blindtext}
\newcommand*{\colorboxed}{}
\def\colorboxed#1#{%
    \colorboxedAux{#1}%
}
\newcommand*{\colorboxedAux}[3]{%
    \begingroup
    \colorlet{cb@saved}{.}%
    \color#1{#2}%
    \boxed{%
        \color{cb@saved}%
        #3%
    }%
    \endgroup
}

\usepackage{pifont}

\usepackage{float}
\usepackage{epsfig}
\usepackage{bm}
\usepackage[matrix,frame,arrow]{xy}
\usepackage[applemac]{inputenc}
\usepackage[T1]{fontenc}
\usepackage{lmodern}
\usepackage[english]{babel}
\usepackage{amsmath}
\usepackage{ae}
\usepackage{amssymb}
\usepackage{color}
\usepackage{graphicx}
\usepackage{bbm}
\usepackage{url}
\usepackage{nomencl}
\usepackage{subfigure}
\usepackage{slashed}

\usepackage{booktabs}

\newcommand{\ket}[1]{|#1\rangle}
\newcommand{\braket}[2]{\langle #1|#2\rangle}
\newcommand{\ketbra}[2]{\left| #1 \rangle \langle #2 \right|}

\newcommand{\sz}{\hat \sigma_z}
\newcommand{\sx}{\hat \sigma_x}
\newcommand{\sm}{\hat \sigma_-}
\renewcommand{\sp}{\hat \sigma_+}

\newcommand{\figref}[1]{\mbox{Fig.~\ref{#1}}}

\renewcommand{\eqref}[1]{\mbox{Eq.~(\ref{#1})}}

\newcommand{\be}{\begin{equation}}
\newcommand{\ee}{\end{equation}}
\newcommand{\bea}{\begin{eqnarray}}
\newcommand{\eea}{\end{eqnarray}}

\usepackage{xr}

\usepackage[colorlinks]{hyperref}

\hypersetup{%
    plainpages=true,
    pdfpagemode={UseOutlines},
    pdfstartview={FitH},
    breaklinks=true,
    hypertexnames=false,
    pageanchor=true,
    colorlinks=true,
    linkcolor={blue},
    citecolor={red},
    urlcolor={blue},
    anchorcolor={black}
}

\newcommand{\beq}{\begin{eqnarray}}
\newcommand{\eeq}{\end{eqnarray}}

\makeatletter

\newcommand{\checknextarg}{\@ifnextchar\bgroup{\gobblenextarg}{and}}
\newcommand{\gobblenextarg}[1]{,\! (\ref{#1})\@ifnextchar\bgroup{\gobblenextarg}{]}}

\usepackage{etoolbox}

\newcounter{eqncount}
\newcommand{\eqsRef}[1]{%
    \setcounter{eqncount}{0}
    \renewcommand*{\do}[1]{\stepcounter{eqncount}}
    \docsvlist{#1}
    Eq%
    \ifnum\value{eqncount}>1
    s
    \fi
    .\nobreakspace
    \def\eqnrefdelim{\def\eqnrefdelim{), (}}%
    \renewcommand{\do}{\eqnrefdelim\ref}%
    \textup{[(}%
    \docsvlist{#1}%
    \textup{)]}%
}

\newcommand{\ac}{\hat a^\dag}
\newcommand{\an}{\hat{a}^{}}
\renewcommand{\sp}{\hat \sigma_+}
\renewcommand{\eqref}[1]{\mbox{Eq.~(\ref{#1})}}

\begin{document}

\title{Atoms in {separated} resonators can jointly absorb a single photon}

    \author{Luigi Garziano}
\affiliation{Theoretical Quantum Physics Laboratory, RIKEN Cluster
for Pioneering Research, Wako-shi, Saitama 351-0198, Japan}

\author{Alessandro Ridolfo}
\affiliation{Dipartimento di Fisica e Astronomia, Universit\`{a}
di Catania, 95123 Catania, Italy}

    \author{Adam Miranowicz}
\affiliation{Theoretical Quantum Physics Laboratory, RIKEN Cluster
for Pioneering Research, Wako-shi, Saitama 351-0198, Japan}
\affiliation{ Faculty of Physics, Adam Mickiewicz University,
PL-61-614
    Poznan, Poland}

    \author{Giuseppe Falci}
\affiliation{Dipartimento di Fisica e Astronomia, Universit\`{a}
di Catania, 95123 Catania, Italy}

    \author{Salvatore Savasta*}
\affiliation{Dipartimento MIFT, Universit\`{a} di Messina, I-98166
Messina, Italy}
\email[corresponding author: ]{ssavasta@unime.it}
    \author{Franco Nori}
\affiliation{Theoretical Quantum Physics Laboratory, RIKEN Cluster
for Pioneering Research, Wako-shi, Saitama 351-0198, Japan}
\affiliation{Department of Physics, University of Michigan, Ann
Arbor, Michigan 48109-1040, USA}



\date{\today}
\begin{abstract}

The coherent nonlinear process where a single photon
simultaneously excites two or more two-level systems (qubits) in a
single-mode resonator has  recently been theoretically predicted.
Here we explore the case where the two qubits are placed in
different resonators in an array of two or three weakly coupled
resonators. Investigating different setups and excitation schemes,
we show that this process can still occur with a probability
approaching one under specific conditions. The obtained results
provide interesting insights into subtle causality issues
underlying the simultaneous excitation processes of qubits placed
in different resonators.

\end{abstract}
 \maketitle

\section{Introduction}

For many decades the possibility of reaching the strong coupling
regime between light and matter has been one of the major topics
of research in atomic physics and quantum optics, driving the
field of cavity quantum electrodynamics (QED). In this regime,
which was first reached in Rydberg atoms interacting with the
electromagnetic field confined in a high-Q
cavity~\cite{Kaluzny1983}, it is possible to observe a coherent
and reversible energy exchange between light and matter, called
vacuum Rabi oscillations, at a coupling rate exceeding the losses
of the system. In 1992, the strong coupling regime was
experimentally achieved~\cite{Thompson1992} also with single atoms
coherently interacting with an optical cavity. Following these
pioneering experiments, this strong regime of light-matter
coupling has been realized in various quantum systems, enabling
tests of fundamental physics and the study of single atom-photon
processes~\cite{Raimond2001}, and leading to important
applications in quantum computation, quantum information
processing, sensing and
metrology~\cite{Haroche2013,Gu2017,Kockum2019book}.

More than two decades after the observation of the strong coupling
regime, the cavity-QED community started investigating the
possibility of accessing a new  non-perturbative light-matter
regime in which the coupling rate can become a significant
fraction of the bare energies of the system. In 2005, it was
predicted ~\cite{Ciuti2005} that this new ultrastrong coupling
(USC) regime could be observed with a planar microcavity photon
mode which is strongly coupled to a semiconductor intersubband
transition in the presence of a two-dimensional electron gas. In
this new USC regime, the rotating-wave approximation (RWA)
employed in the standard Jaynes-Cummings
model~\cite{Jaynes1963,Shore1993}, which has been a workhorse of
quantum optics in the weak and strong coupling regimes, cannot be
safely applied anymore \cite{Kockum2019,Forn-Diaz2018}. Indeed, it has been shown that the
counter-rotating terms in the system Hamiltonian become
relevant (see, e.g., Ref.~\cite{DeLiberato2007, Gambino2014,Forn-Diaz2016,Zueco2019}), giving
rise to a wide variety of novel and unexpected physical phenomena
~\cite{Ashhab2010,Garziano2013,Stassi2013,Huang2014,DeLiberato2014,
Garziano2015,Stassi2016,Garziano2016,
DeLiberato2017,Stassi2017,Kockum2017,Falci2017,Bin2019,Falci2019}.

A proper description of the USC-regime physics requires to solve
some fundamental theoretical  issues, such as the failure of the
usual normal-order correlation functions to describe the correct
output photon emission
rate~\cite{Ciuti2006,Ridolfo2012,LeBoite2016,DiStefano2017a}, some
unphysical predictions of the standard master equation
approach~\cite{Beaudoin2011,Bamba2012,Settineri2018}, and gauge
ambiguities in the quantum Rabi and Dicke
models~\cite{DiStefano2018,Stokes2018,DeBernardis2018}. Besides
the vast phenomenology that has been predicted to be observable in
this new light-matter regime, the interest  has also been fostered
by the experimental realization of USC in several physical
systems, including superconducting quantum
circuits~\cite{Johansson2009,
Bourassa2009,Forn-Diaz2010,Niemczyk2010,
Baust2016,Bosman2017a,Bosman2017,Yoshihara2017a,
Langford2017,Chen2017,Yoshihara2017,Gheeraert2018,Semba2018},
intersubband polaritons in microcavity-embedded doped quantum
wells~\cite{Ciuti2005,Gunter2009}, and other hybrid
cavity-QED systems, such as Landau
polaritons~\cite{Hagenmuller2010,Hagenmuller2012,Bayer2017,Jeannin2019,Keller2020},
microcavity exciton
polaritons~\cite{Jouy2011,Gubbin2014,Kono2018,Li2018,Hou2019},
magnons in microwave cavities~\cite{Goryachev2014,Zhang2015} and
organic molecules~\cite{Schwartz2011,
Kena-Cohen2013,Gambino2014,Mazzeo2014,
Galego2015,Bennett2016,Kowalewski2016,
George2016a,Herrera2016,Ebbesen2016,Galego2017,
Kolaric2018,Genco2018}. Among  unique physical effects of the USC
regime, there are those related to the hybridization of the ground
state of the quantum Rabi Hamiltonian~\cite{Ciuti2005,
DeLiberato2007,Garziano2013,Cirio2016,
DeLiberato2017,DiStefano2017,SanchezBurillo2019}. Such a ground
state now contains \textit{virtual} excitations that can be
released only by applying a time-dependent perturbation to the
system. Moreover, the USC regime  opens the possibility of
observing higher-order processes and nonlinear optics  with
two-level systems and  virtual photons~\cite{Stassi2017,
Kockum2017,Kockum2017a}, symmetry breaking and Higgs
mechanism~\cite{Garziano2014}, multiphoton quantum Rabi
oscillations~\cite{Garziano2015}, and even more counterintuitive
phenomena like the emission of bunched light from individual
qubits~\cite{Garziano2017}.

One of the most interesting nonlinear optical effects predicted in
the USC regime consists of the  simultaneously excitation of  two
or more spatially separated atoms by a single
photon~\cite{Garziano2016,Zhao2017,Wang2017,Macri2020}. This last
puzzling result, which has been studied in a quantum system
constituted by two qubit ultrastrongly coupled to a single-mode
resonator, provides new insights into the various quantum aspects
of the interaction between light and matter and can find useful
applications for the development of novel quantum technologies.
Although this effect clearly demonstrates a relevant role of the
counter-rotating terms and virtual processes in the USC regime,
the single-mode approach does not allow to fully understand some
subtle causality issues underlying this process, since the
resonator mode is completely delocalized along the cavity.
Specifically, a drawback of  this simplified description of the
electromagnetic field is that {\em any information about the
spatial separation between the two atoms is lost}. Hence, the
question arises if it is possible to observe this effect in the
presence of natural or artificial atoms which are {\em actually}
spatially separated. Here we provide a positive answer to this
question, even though further work will be required for a full
understanding of the impact of the spatial separation of the atoms
on the joint absorption and emission of single photons. Moreover,
it has been recently pointed out~\cite{SanchezMunoz2018} that  the
description of a cavity-QED system in terms of the single-mode
quantum Rabi model in the USC and deep-strong coupling (DSC)
regimes can lead to the violation of relativistic causality, and a
multi-mode version of the quantum Rabi model is required in order
to {fully capture the propagation properties of the light field necessary
to comply with causality.}
{Here, although we do not adopt a multimode approach, we still assume the atoms to be spatially separated. Specifically, we introduce a simplified description of spatial separation, considering the two atoms embedded in different single-mode resonators. For example an array of $N$ nearest-neighbour single-mode coupled resonators corresponds to a system with $N$ distinct sites, like in tight-binding models in solid state physics, where propagation effects are taken into account. In these models the propagation speed depends on the interaction rate $J$ between  nearest neighbours cavities. For example the complete transfer of a photon from one resonator to the adjacent one requires a time $T = \pi / (2J)$.}

In the present work, we show that the simultaneous excitation
process of two qubits by a single cavity photon, as described in
Ref.~\cite{Garziano2016}, can take place also in a cavity-array
system of two or three cavities, where the qubits are placed in
different resonators and ultrastrongly interact with them. We
observe that this effect can be achieved by probing the system via
two different excitation mechanisms: (i) by exciting one of the
normal modes of the coupled-cavity array or (ii) by selectively
exciting only a single cavity. The substantial difference between
the two cases is that, while the first case corresponds to the
interaction of two qubits with a delocalized field and  leads to a
deterministic simultaneous excitation process as in
Ref.~\cite{Garziano2016}, the second case constitutes one of the
simplest examples of a localized system in which the excitation is
initially stored in a single resonator. This makes this effect
even more counterintuitive, since the excitation is continuously
transferred between the nearest-neighbor resonators, while  the
effect requires  both atoms to feel the photon field at the same
time in order to take place. In both cases, we study the temporal
evolution of the system within both  theoretical and numerical
approaches, providing a clear and physically intuitive description
of the propagation and causality mechanisms underlying this
simultaneous excitation phenomena. The process described here
could be experimentally realized in state-of-the-art circuit QED
systems. Moreover, these effects can  find useful applications in
quantum information processing and quantum communication
protocols, where reliable and controllable entanglement  between
distant qubits in a quantum network is of fundamental importance \cite{Leung2019,Kato2019}.

\section{Models and Results} \label{sec2}

Here we study a quantum system consisting of an array of $ N $
weakly coupled single-mode  resonators, each of them interacting
with a two-level atom (e.g., a superconducting qubit). The total
Hamiltonian of the system can be written as~\cite{Garziano2016}
(hereafter, $ \hbar=1 $):

 \begin{equation} \label{1}
 \hat{H} = \hat{H}_{ q} + \hat{H}_{ c} + \hat{H}_{\rm cc}+ \hat{H}_{\rm qc}\, ,
 \end{equation}
where  $H_{c}=\sum_{n=1}^{N} \omega_{c}^{(n)} \ac_n \an_n $ and
$H_{q}=\sum_{n=1}^{N} \omega_{q}^{(n)} \sp^{(n)} \sm^{(n)} $
describe, respectively, the qubit and cavity Hamiltonians in the
absence of interaction, $ \ac_n (\an_n)$  is the bosonic creation
(annihilation) operator for the $ n $th resonator mode with
frequency $ \omega_{c}^{(n)} $,  and $\sp^{(n)}(\sm^{(n)})  $ are
the raising (lowering) operators for the $ n $th qubit with
transition frequency $ \omega_{q}^{(n)} $. The cavity-cavity
interaction Hamiltonian is given by
\be \label{2} H_{\rm cc}= J
\sum_{n=1}^{N-1} \hat{a}^\dag_{n}\hat{a}_{n+1}\, , \ee
where $J$ is the  next-neighbour
hopping rate. {We are assuming that  the coupling strength $J$ between the two
cavities is weak, thus we  can apply the RWA to the cavity-cavity
interaction Hamiltonian $ \hat{H}_{\rm cc} $}.  Finally, the last term of \eqref{1}, describing the
interaction between the qubits and the cavity modes, reads
 \be \label{3}
 H_{\rm qc}=\sum_{n=1}^{N} g_{n}\, \hat{X}_{n}
 [\cos(\theta_{n})\sx^{(n)} + \sin(\theta_{n})\sz^{(n)}]   \, ,
 \ee
where $ g_n\equiv|g_n|e^{i \varphi_n} $ denotes the coupling rate
of the $ n $th qubit to the corresponding cavity field $\hat{X}_{n}\equiv \ac_n+ \an_n $, {the angles
$ \theta_{n} $  parametrize  the relative contribution of the
transverse and longitudinal couplings, while  $\sx^{(n)}$ and
$\sz^{(n)}  $  are the Pauli matrices for the qubits. In circuit
QED systems, the angle $ \theta_n $ and the transition frequency $
\omega_{q}^{(n)} $ can be continuously tuned by changing the
magnetic field  externally applied to, e.g., a flux qubit (see,
e.g., Ref.~\cite{Niemczyk2010}). An important feature of the
interaction Hamiltonian is that it contains terms  that do
\textit{not} conserve the total number of excitations.
Specifically, the \textit{transverse}  coupling $ \propto  \sx $
contains terms like $ \an \sm $ and $ \ac \sp $ which create or
annihilate \textit{two} excitation simultaneously; whereas the $
\sz $-coupling changes the resonator photon number by one, while
leaving the number of qubit excitations unchanged. Notice that the
parity of qubit $n$ is conserved only for $\theta_n = 0$,
corresponding, for a flux qubit, to a zero external flux
offset~\cite{Niemczyk2010}. The terms in the total Hamiltonian
which do not conserve the number of excitations in the system can
be safely neglected in the weak-coupling regime, where the rotating-wave approximation (RWA) is valid. However, these terms become relevant for
systems entering the USC regime, where the coupling strength $
g_{n}  $ reaches an appreciable fraction of the unperturbed
frequencies (here, $ \omega_{c}^{(n)} $ and $ \omega_{q}^{(n)} $)
of the bare systems. One of the most interesting consequences of
the presence of these counter-rotating terms is the possibility to
coherently couple quantum states with  \textit{different} numbers
of excitations. These unconventional  couplings determine new
intriguing physical processes as, for example, multiphoton Rabi
oscillations~\cite{Garziano2015} and the possibility to excite two
or more spatially separated atoms with a single
photon~\cite{Garziano2016}. Here, we  investigate the latter process, considering a fundamentally   different setup, consisting
of weakly coupled cavity arrays, where each cavity ultrastrongly
interacts with a single qubit. Our results show that, owing to the
ultrastrong light-matter interaction and the parity-symmetry
breaking, it is still possible to simultaneously excite two
qubits, even if they are placed in different cavities.}

\subsection{Two cavities, two qubits and one photon}

\begin{figure}[!h]
 \centering
\includegraphics[scale=0.45]{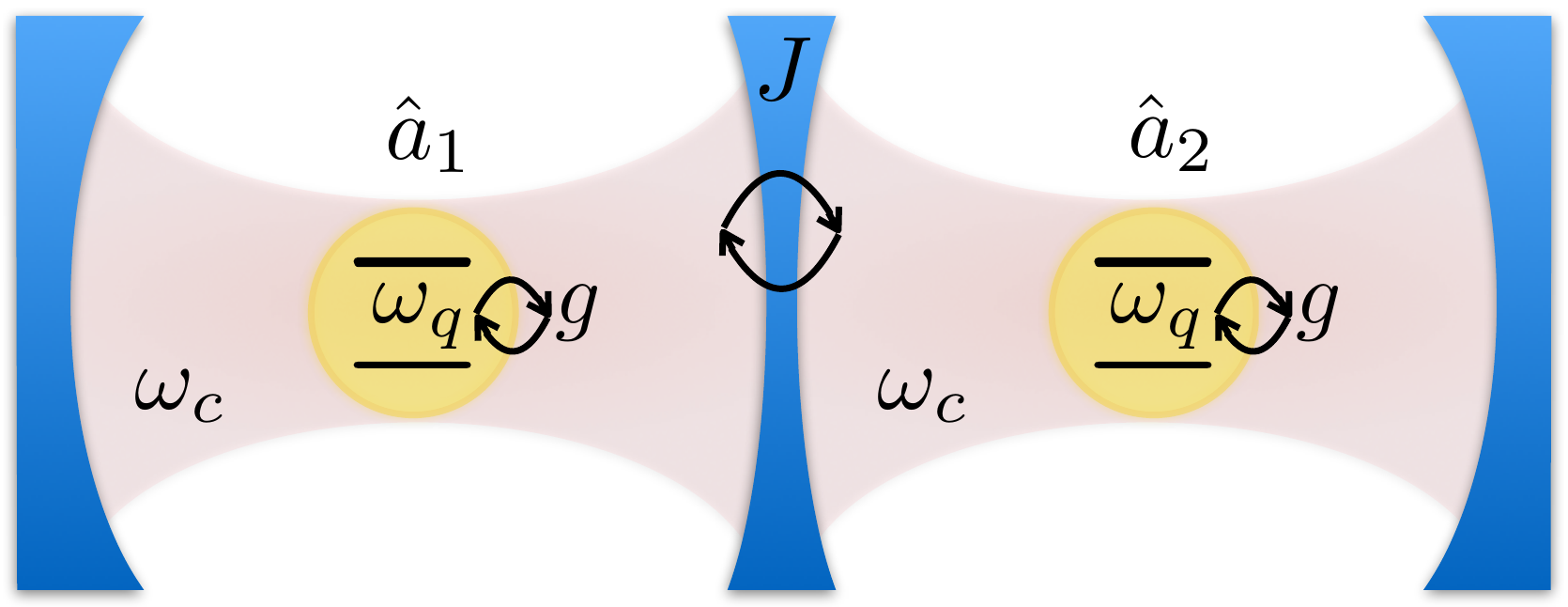}
\caption{Sketch of the system. Two identical spatially separated
optical resonators,  with resonance frequency $\omega_{c}$, are
weakly coupled together, each one of them ultrastrongly
interacting with a single two-level system (qubit) with transition
frequency $\omega_{q}$. The photon hopping rate $J$ between the
two  resonators and the light-matter coupling strength $ g$ are
indicated.} \label{fig1}
\end{figure}

We first focus on the study of an array of two identical
cavity-qubit systems $ \bigl(
\omega_c^{(1)}=\omega_c^{(2)}=\omega_{c}\, ;
\,\omega_q^{(1)}=\omega_q^{(2)}=\omega_{q}\, ;
\,\theta_{1}=\theta_{2}=\theta \bigr)  $, where the two cavities
are weakly coupled together and each one of them ultrastrongly
interacts with a single qubit (see \figref{fig1}). In this case,
the Hamiltonian of \eqref{1} becomes
\be \label{h2cav}
 \hat{H} =\sum_{n=1}^2\left[  \omega_{c} \,\ac_n \an_n +  \omega_{q}\, \sp^{(n)} \sm^{(n)} +\left| g\right|e^{i \varphi_n}  \hat{X}_{n} \left(   \cos\theta\, \sx^{(n)} + \sin \theta\,\sz^{(n)}  \right)  \right] + J  \left(\an_1 \ac_2 + \ac_1 \an_2  \right)  \,,
\ee
The Hamiltonian in
\eqref{h2cav}  can be conveniently rewritten in terms of the
bosonic  symmetric and antisymmetric normal modes, also referred
to as supermodes defined via the operators $
\hat{a}^{}_{S(A)}=(\hat{a}_{1}\pm \hat{a}_{2})/\sqrt{2} $, which
diagonalise the Hamiltonian
 \be
 \hat{H}_C=\sum\limits_{n=1}^2 \omega_{c} \,\ac_n \an_n  + J  \left(\an_1 \ac_2 + \ac_1 \an_2  \right)\,,
 \ee
 describing  two weakly coupled harmonic oscillators.

 In this case, we obtain
\bea \label{h2cavnew}\nonumber
 \hat{\mathcal{H}}&=&\omega_{A} \,\hat{a}^\dag_{A} \hat{a}_{A}^{} +\omega_{S} \,\hat{a}_{S}^{\dag}  \hat{a}_{S}^{}+
   \omega_{q}\sum_{n=1}^2  \sp^{(n)} \sm^{(n)} \\
&+& | g  | \left[\hat{X}_{A} \left(\cos\theta \, \hat{\Phi
}_{x}^{-} +  \sin \theta\, \hat{\Phi }_{z}^{-} \right)+\hat{X}_{S}
\left(\cos\theta \, \hat{\Phi }_{x}^{+} +  \sin \theta\, \hat{\Phi
}_{z}^{+}  \right)  \right]\,, \eea where $ \hat{X}_{S(A)}\equiv
\hat{a}^\dag_{S(A)} + \hat{a}_{S(A)}^{}$ and $\hat{\Phi
}_{x(z)}^{\pm}\equiv \left( e^{i
\varphi_1}\,\hat{\sigma}_{x(z)}^{(1)} \pm e^{i \varphi_2}\,
\hat{\sigma}_{x(z)}^{(2)}\right)/\sqrt{2}$. The Hamiltonian in
\eqref{h2cavnew}  describes  two bosonic modes, one symmetric and
one antisymmetric with corresponding frequencies $
\omega_{S}=\omega_{c}+ J$ and $ \omega_{A}=\omega_{c}- J$, both
interacting with two qubits.
We now diagonalise numerically $\hat{\mathcal{H}}$, indicating the
resulting energy eigenvalues and eigenstates as $ \omega_{i} $ and
$ \ket{E_i} $, with $ i = 0,1, \dots , $. We  label the states such
that $ \omega_{k}>\omega_{j} $ for $ k > j $. In our analysis, we
use  the notation
$\ket{\mathcal{N}_A,\mathcal{N}_S,q_1,q_2}=\ket{\mathcal{N}_A}\bigotimes
\ket{\mathcal{N}_S}\bigotimes\ket{q_1}\bigotimes \ket{q_2} $ for
the eigenstates $ \ket{E_i} $, where $q=\left \{g,e\right \}$
denotes the qubit ground or excited states, respectively, and
$\ket{\mathcal{N}_{S(A)}}=\left \{\ket{0},\ket{1},\ket{2},\dots
\right \} $ represents the Fock states with a photon occupation
number $ \mathcal{N}$ in the symmetric (antisymmetric) normal
mode.

\begin{figure}[!ht]
\includegraphics[scale=0.45]{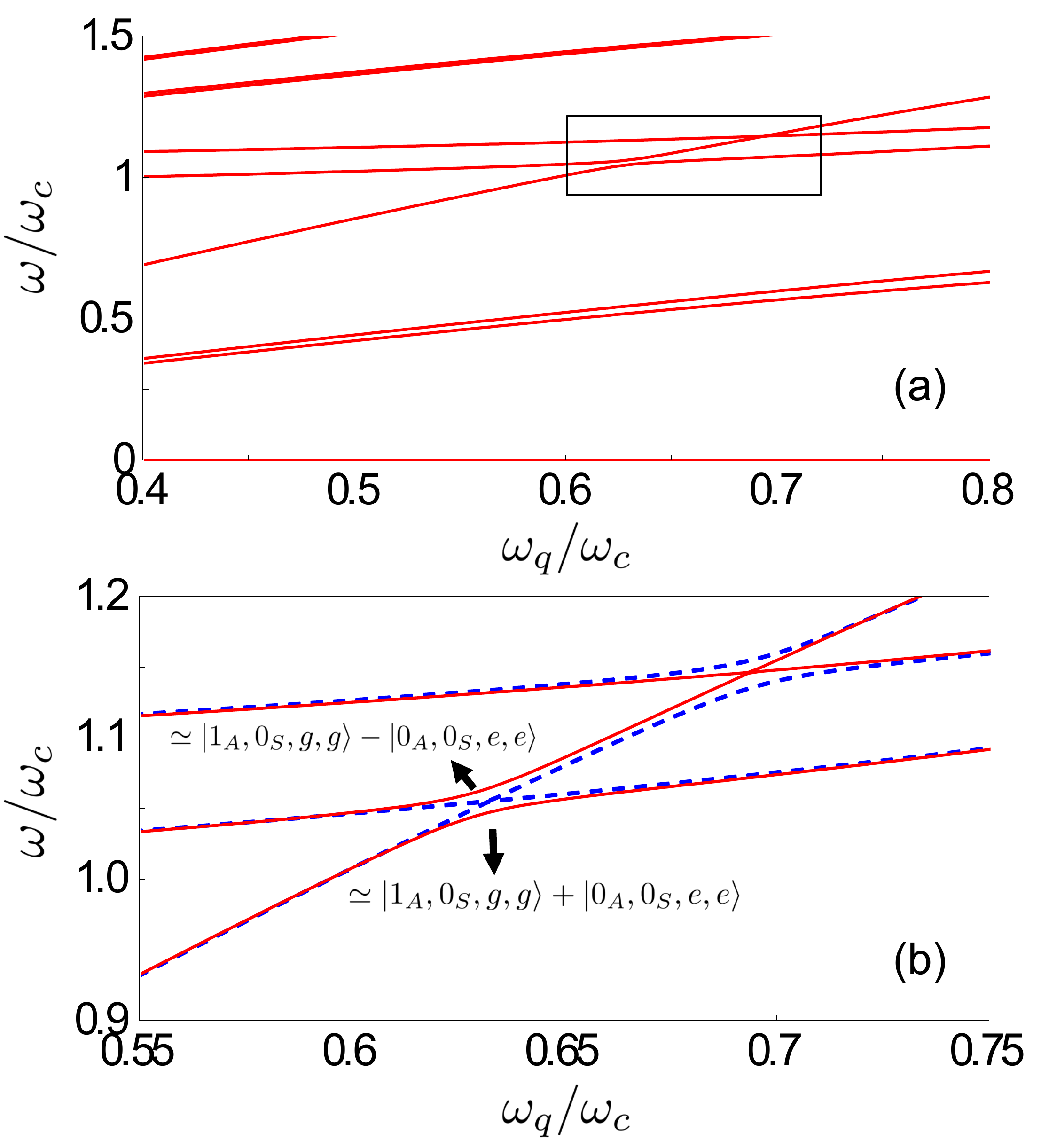}
\caption{(a) Energy differences $
\omega_{i0}=\omega_{i}-\omega_{0}$ for the lowest-energy dressed
states of  $\hat{\mathcal{H}} $ as a function of the normalized
qubit frequency $ \omega_{q}/\omega_{c}$ (which can be
experimentally tuned by changing the external flux bias acting on
the qubits). We consider a normalized coupling rate $ \eta
\equiv|g|/\omega_{c}=0.3 $ between the qubit and the resonators,
while the normalized photon hopping rate between the two
resonators is $ J/\omega_{c}=0.05 $. The phases for the
cavity-qubit coupling strengths are set to $\varphi_{1}=0  $ and
$\varphi_{2}=\pi$, respectively, and the longitudinal interaction
coupling term is included by considering a mixing angle $ \theta=
\pi/6 $. (b) Enlarged view of the inset in (a). When the
cavity-qubit coupling strengths have opposite phases
($\varphi_{1}=0,\varphi_{2}=\pi $), the avoided-level crossing
(red solid curves) results from the coupling between the states
$\ket{1_A,0_S,g,g}$ and $\ket{0_A,0_{S},e,e}$ due to the presence
of counter-rotating terms in the system Hamiltonian. The energy
splitting reaches its minimum at $\omega_{q}  \simeq \omega_{A}/2
$. The crossing between the states $\ket{0_A,1_S,g,g}$ and
$\ket{0_A,0_{S},e,e}$ at $\omega_{q}  \simeq \omega_{S}/2 $
indicates that the qubits do not couple with the symmetric normal
mode. The complementary result is obtained when the coupling
strengths have the same phase (blue dashed curves). In this case,
this splitting at $\omega_{q} \simeq \omega_{A}/2$ disappears,
while the energy spectrum  displays an avoided-level crossing
around $ \omega_{q} \simeq \omega_{S}/2  $, arising from the
coherent coupling between the states $\ket{0_A,1_{S},g,g}$ and $
\ket{0_A,0_{S},e,e}$.} \label{fig2}
\end{figure}

We set the normalized hopping rate and the light-matter coupling
strength as $ J/\omega_{\rm c}=0.05 $ and  $ \eta\equiv
|g|/\omega_{\rm c}=0.3 $, respectively. Moreover, we set the
phases for the cavity-qubit coupling strengths to $\varphi_{1}=0 $
and $\varphi_{2}=\pi$, respectively. This specific phase
difference does not affect the dynamics, however it will play an
important role for the case of three coupled cavities (see
Sect.~\ref{sec2B}). We also consider a mixing angle $ \theta=
\pi/6 $, such that both the longitudinal and transverse
contributions to the interaction have comparable values. Figure
\ref{fig2}(a) shows the energy differences $
\omega_{i0}=\omega_{i}-\omega_{0}$ for the lowest-energy states as
a function of the normalized qubit frequency $
\omega_{q}/\omega_{c}$, which can be experimentally tuned by
changing the external magnetic flux acting on the qubit. We
observe that, when $\omega_{q} \simeq \omega_{A}/2  $, the
spectrum displays an avoided-level crossing  between the states $
\ket{E_3} $ and $ \ket{E_4} $. {It is worth noticing that the
resonance condition is quite different from the expected one, i.e.,
$\omega_{q}  = \omega_{A}/2 $, because $\omega_q$ and $\omega_A$
are bare resonance frequencies of the matter and light components.
The actual physical frequencies are significantly dressed by the
interaction (see, e.g.,~\cite{Zueco2009,Garziano2015,
Garziano2016, Stassi2017, DiStefano2017}). Note that, just outside this avoided-level crossing region, one level remains flat
as a function of the qubit frequency with energy $\omega \approx
\omega_{A}$, while the other shows a linear behaviour with $\omega
\approx  2 \omega_{q}$.  The origin of this splitting is due to the
hybridization of the states $\ket{1_A,0_S,g,g}$ and
$\ket{0_A,0_{S},e,e}$.}
 When the splitting is at its minimum, these states are well approximated by the superposition states
 \be
 \ket{E_{3(4)}}=\left( \ket{1_A,0_S,g,g}\pm\ket{0_A,0_{S},e,e}\right)/\sqrt{2}\, .
 \ee
 The (numerically calculated) normalized minimum splitting has a value $ 2 \Omega_{\rm eff}/\omega_{c}=16 \times 10^{-3}$, where $ \Omega_{\rm eff} $ is the effective coupling rate between two qubits and one photon. It is important to observe that the coherent coupling between these two states  would not be allowed within the RWA, since they have a different number of excitations.
Moreover, as the excitation number difference between the two
states is odd,  parity-symmetry breaking ($ \theta \ne 0 $) is
required in order to observe this splitting. As reported in
Ref.~\cite{Garziano2016}, the effective coupling between the
states $ \ket{E_3} $ and $ \ket{E_4} $ can be analytically
described by an effective Hamiltonian. Moreover, this coherent
coupling is not direct, but can only occur  via virtual
transitions which are enabled by the counter-rotating terms in $
\hat{\mathcal{H}}$. In this way,  the initial state
$\ket{1_A,0_{S},g,g}$  evolves to virtual intermediate states that
eventually do not  conserve the energy, but it finally evolves to
a real energy-conserving state, i.e., $ \ket{0_A,0_{S},e,e} $. It
is interesting to observe that, for  $ \omega_{q} \simeq
\omega_{S}/2  $,  the energy spectrum displays a  crossing
between the levels $\ket{0_A,1_{S},g,g}$ and $ \ket{0_A,0_{S},e,e}
$, showing that the two qubits do not interact with the symmetric
normal mode of the coupled cavities. Indeed, if the coupling
strengths $g_{1}  $ and  $g_{2}  $  have opposite signs it can be
shown that all the possible intermediate virtual transitions for
the process $\ket{0_A,1_{S},g,g}\to \ket{0_A,0_{S},e,e} $, induced
by the interaction term proportional to $ \hat{X}_{S} $, lead to
an intermediate state proportional to $
\ketbra{e}{e}-\ketbra{e}{e}$, thus  giving a vanishing
contribution.

Figure \ref{fig2}(b) shows  the comparison of the avoided-level
crossing behavior for the cases $\varphi_{1}=0,\varphi_{2}=\pi  $
(solid red curves) and $\varphi_{1}=\varphi_{2}=0$ (dashed blue
curves). It can be observed that the two choices lead to
complementary results. Indeed, when the two coupling strengths
have same signs, the splitting at $\omega_{q} \simeq \omega_{A}/2$
disappears, while we observe the presence of an avoided-level
crossing around $ \omega_{q} \simeq \omega_{S}/2  $ arising from
the coherent coupling between the states $\ket{0_A,1_{S},g,g}$ and
$ \ket{0_A,0_{S},e,e}$. This result shows that, depending on the
relative signs of the coupling strengths, for $ N=2 $  the
simultaneous excitation of two qubits placed in different
resonators can be achieved by coupling the qubits either to the
symmetric or antisymmetric normal mode.

In order to fully understand and characterise this process, we fix
the qubit frequency  at the value where the splitting, as shown in
Fig.~\ref{fig2}(a), between the energy levels corresponding to the
eigenstates $ \ket{E_{3}} $ and $ \ket{E_{4}} $  is minimum and
consider the system initially prepared in the one-photon state
$\ket{1_A}\equiv\ket{1_A,0_{S},g,g}$. As we will see later, the
preparation of this state can be experimentally achieved by
sending an appropriate electromagnetic Gaussian pulse to the first
cavity.

\begin{figure}[h]
    \includegraphics[scale=0.45]{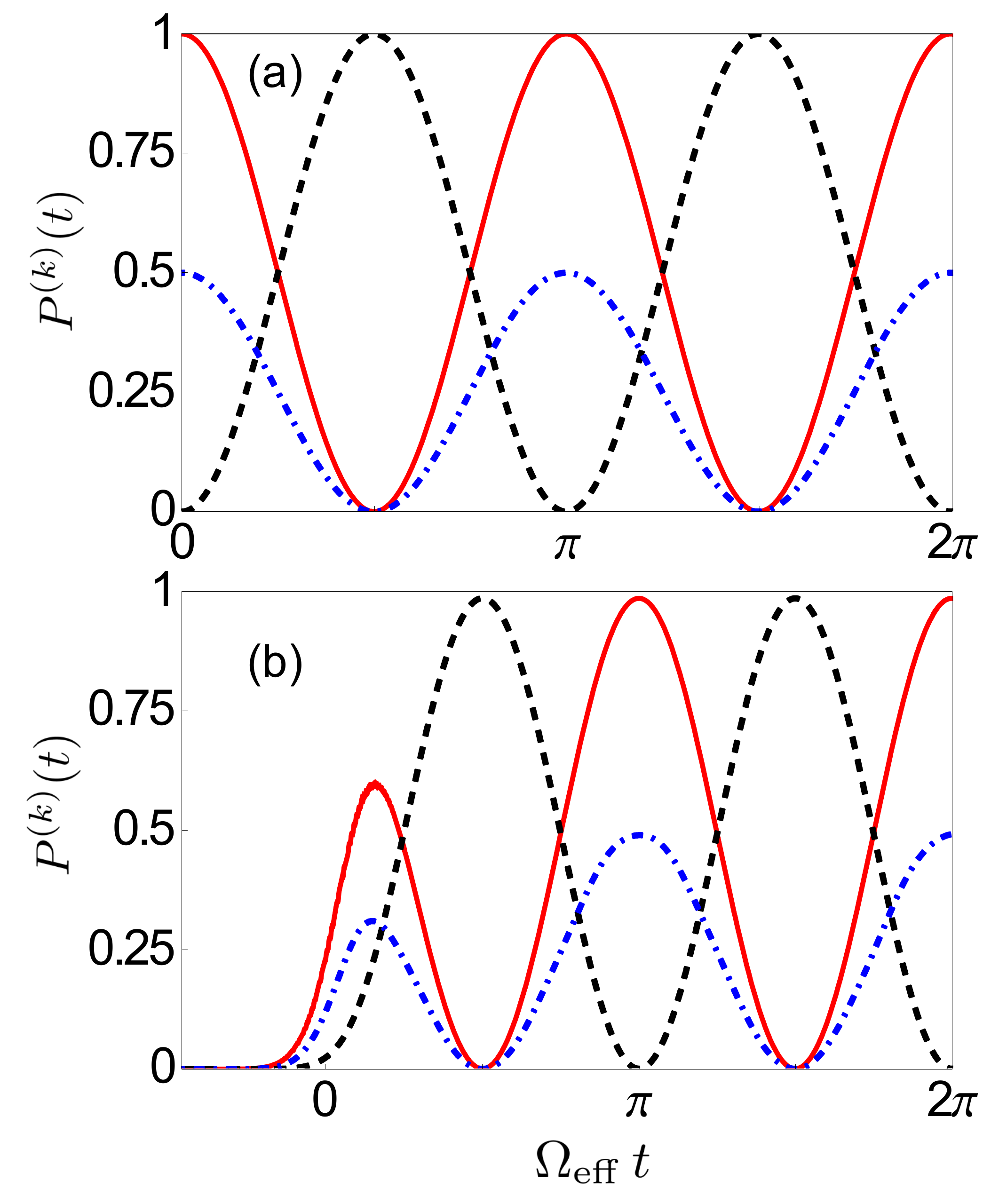}
    \caption{{(a) Time evolution of the occupation probabilities $P^{(k)}(t)\equiv \langle \hat{\mathcal{P}}_{k} \rangle $, with $\hat{\mathcal{P}}_{k}= \ketbra{k}{k}  $, for the single-photon states $\ket{1_A}$ (red solid curve) and $ \ket{1_1} $ (blue dot-dashed curve), together with the probability  $ \mathcal{P}^{(ee)} (t)$ of having both qubits simultaneously excited (black dashed curve) for the system initially prepared in the state $\ket{1_A}\equiv\ket{1_A,0_{S},g,g}$. Vacuum Rabi oscillations showing a reversible excitation exchange process between the qubits and the resonators are clearly visible. The joint absorption of an antisymmetric cavity photon by the two qubits is achieved after a Rabi half  period $ \Omega_{\rm eff}\,t=\pi/2$, with the excitation probability  $ \mathcal{P}^{(ee)} $ approaching one, even if they are placed in different resonators. Here, the effects of dissipation have  not been included. (b) Temporal evolution of the same occupation probabilities $P^{(k)}(t)  $ considered in (a), but after the arrival of a narrow Gaussian pulse  exciting the first cavity when the system is initially prepared in its ground state. The amplitude and the central frequency of the pulse are $ A/\omega_c = 4.2\times 10^{-2} $  and  $\omega_d=(\omega_{30}+\omega_{40})/2 $, respectively.}}
    \label{fig3}
\end{figure}

Figure \ref{fig3}(a) displays the numerically calculated time
evolution of the occupation probabilities $P^{(k)}(t)\equiv
\langle \hat{\mathcal{P}}_{k} \rangle $, with
$\hat{\mathcal{P}}_{k}= \ketbra{k}{k}  $, for the  one-photon
states $\ket{1_A}$ (i.e., one photon in the antisymmetric normal
mode) and $ \ket{1_1} $ (one photon in the first cavity), together
with the probability  $ \mathcal{P}^{(ee)} (t)$ of having both
qubits simultaneously excited. In terms of the dressed energy
eigenstates of the system, these states can be expressed,
respectively, as
 \bea
 &\ket{1_A}&=\left( \ket{E_3}+\ket{E_4}\right) /\sqrt{2}\, ,\\   &\ket{1_1}&=\left(\ket{E_3}+\ket{E_4}+\sqrt{2} \,\ket{E_5}\right)/2\, ,\\
 &\ket{e,e}&=\left( \ket{E_3}-\ket{E_4}\right) /\sqrt{2}\,,
 \eea
where, for concision, we omitted in the last equation the photonic
states, which are intended to be in the ground state. Here, $|e,e
\rangle$ stands for $|0_A, 0_S,e,e \rangle$. As expected, since
$\ket{1_A}=\left(\ket{1_1,0_2}-\ket{0_1,1_2} \right)|g,g
\rangle/\sqrt{2} $,  at the initial instant of time $
\mathcal{P}^{( 1_{ A})} (0)=1$, while $ \mathcal{P}^{( 1_1)}
(0)=1/2$. As time evolves, vacuum Rabi oscillations, showing the
reversible excitation exchange process between the qubits and the
resonators, are clearly visible. Specifically, we observe that,
after a Rabi half  period $ \Omega_{\rm eff}\,t=\pi/2$, one photon
in the antisymmetric cavity mode  is jointly absorbed by the two
qubits, even if they are placed in different resonators. Moreover,
the excitation probability  $ \mathcal{P}^{(ee)} $ approaches one,
showing that the multiatom absorption of a single photon can
essentially be deterministic. Notice that, in order to provide a
clearer description  of  this counter-intuitive excitation
mechanism, the effects of dissipation have  not been taken into
account. This  approximation becomes experimentally reasonable
when the system loss rates are smaller than the frequency
splitting between the levels involved at the avoided-level
crossing, so that the first Rabi cycles are almost not affected by
dissipation. A similar oscillating dynamics can be obtained for
the system initially prepared in the state $| e,e \rangle$. In
this case, the two qubits will jointly and coherently release
their energy to the cavity. The time evolution of the system will
be as shown in \figref{fig3}(a), but with the initial time $t=
\pi/ (2\Omega_{\rm eff})$.

As mentioned before, instead of starting from the ideal initial
state $ \ket{1_A} $, we now consider a more realistic case where
the system is initially in its ground state
$\ket{E_0}=\ket{0_A,0_S,g,g}$ and study a direct excitation of the
first cavity by an electromagnetic Gaussian pulse. The
corresponding driving Hamiltonian is
 \be \label{pulse1}
 \hat{H}_{d}=\mathcal{E}(t)\cos(\omega_d \,t)\hat{X}_{1} \, ,
 \ee where $\hat{X}_1= \hat{X}_{S}+\hat{X}_{A} $ and \be
 \mathcal{E}(t)=A \exp \left[ - (t-t_0)^2/(2 \tau^2) \right] /
 (\tau \sqrt{2} \pi)\,,
 \ee
with $ A $ and $ \tau $ the amplitude and the standard deviation
of the Gaussian pulse, respectively. The central frequency of the
pulse has been chosen to be in the middle of the two split
transition energies $\omega_d=(\omega_{30}+\omega_{40})/2 $. The
pulse bandwidth must be sufficiently narrow in order to  ensure
that only the states $ \ket{E_{3}} $ and $ \ket{E_{4}} $ are
excited, so that the pulse can directly excite the state $
\ket{1_A}=\left( \ket{E_3}+\ket{E_4}\right) /\sqrt{2}$, and the
symmetric mode $|1_S \rangle$ is not excited. {This corresponds to
a pulse duration $\sim \tau$ significantly larger than the transfer time $\sim\pi/(2J)$ of photons from one cavity to the other. In this way, even if the system is fed through a single cavity only, both cavities are actually excited simultaneously
without any causality issue, because the excitation time $\tau$ is
larger than the transfer time $\sim \pi/(2 J)$. With this
excitation scheme, the resulting dynamics is very similar to the
case of two atoms in a single cavity~\cite{Garziano2016}.}

Figure~\ref{3}(b) shows the dynamics of the  occupation
probabilities $ \mathcal{P}^{( 1_{ A})}$,  $ \mathcal{P}^{( 1_{
1})}$,  and  $ \mathcal{P}^{( ee)}$ after the arrival of the $ \pi
$-like Gaussian pulse initially exciting the first cavity
described by the Hamiltonian in \eqref{pulse1}. We observe that,
since the pulse time width is not much narrower than the Rabi
period, after the arrival of the pulse, the antisymmetric normal
mode is not completely populated and the excitation is partially
transferred to the qubits. Therefore, the first peak of the
antisymmetric normal mode occupation probability in
\figref{fig3}(b) is slightly lower than the second one.  Once the
antisymmetric mode is completely populated, the dynamics of vacuum
Rabi oscillations, showing the reversible excitation exchange
between one photon in the antisymmetric normal mode and the two
qubits, is the same as in \figref{fig3}(a). It is also worth noticing
that, in the absence of any system nonlinearity as, e.g., in the
case of a system empty (without atoms) resonators, coherent
excitation, as described by the Hamiltonian in \eqref{pulse1},
would give rise to a coherent intra-cavity field, not to a single-photon Fock state. However, the very strong interaction of the
cavity array with the two atoms induces an anharmonicity to the
level structure which is able to prevent the resonant excitation
of higher-photon states (photon-blockade, see,
e.g.,~\cite{Birnbaum2005,Hoffman2011}). {In the case of lower
atom-cavity coupling strengths, the photonic system could be
complemented by additional Kerr nonlinearities, which are able to induce photon blockade~\cite{Garziano2016}.}

\begin{figure}[!ht]
\centering
\includegraphics[scale=0.45]{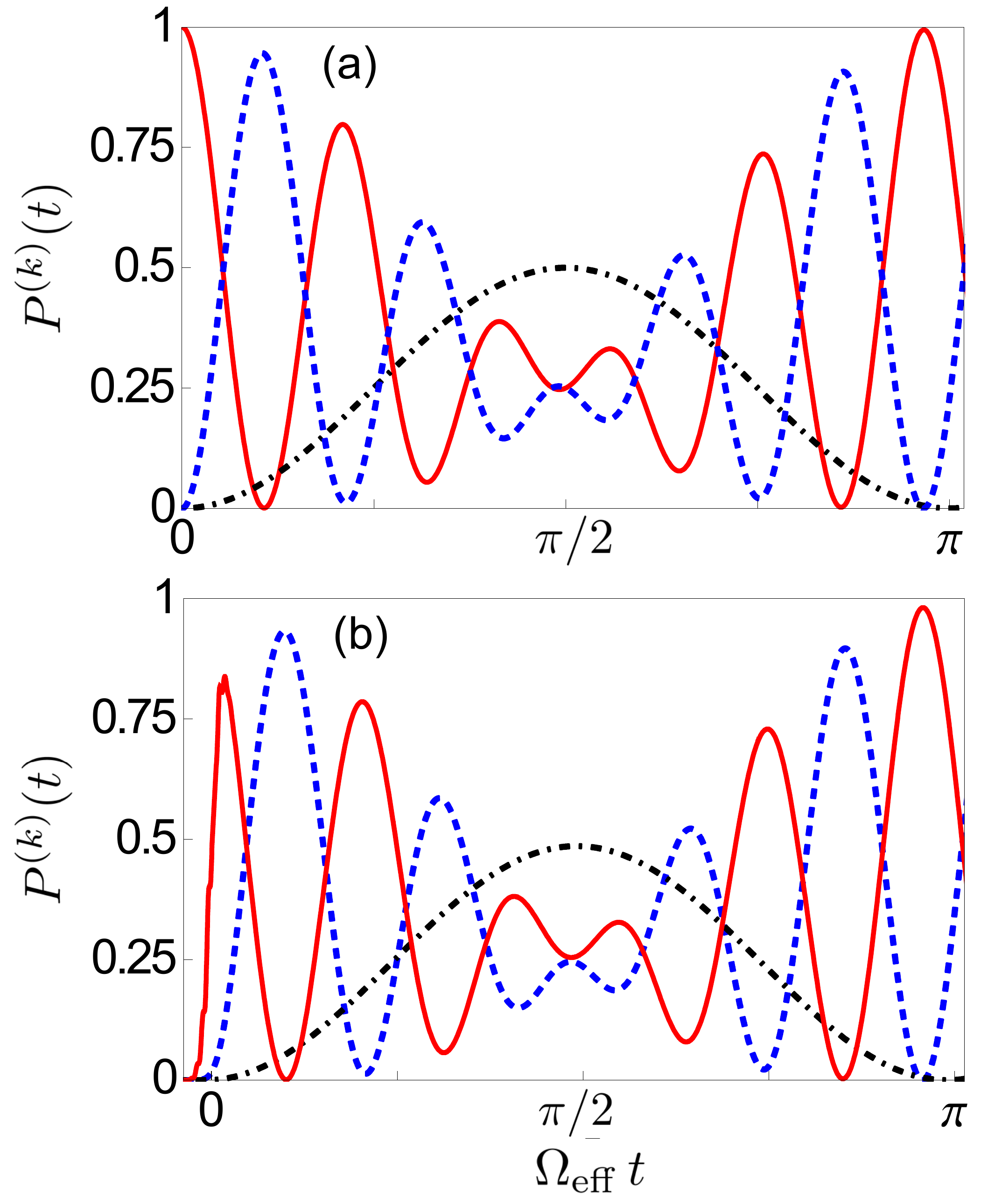}
\caption{{(a) Time evolution of the occupation probabilities $
\mathcal{P}^{( 1_1)} (t)$ (red solid curve), $ \mathcal{P}^{(
1_2)} (t)$ (blue dotted curve), and $ \mathcal{P}^{( ee)} (t)$
(black dot-dashed curve) with the system initially prepared in the
one-photon state $\ket{\psi_1}$ given in \eqref{E1}. As the time
evolves, the expected photon hopping between the two
weakly-coupled cavities is accompanied by the simultaneous
excitation of the two qubits by a single-cavity photon. The
maximum probability $ \mathcal{P}^{( ee)}=1/2$ is achieved at $
t=\pi/(2\Omega_{\rm eff})$. The corresponding values $
\mathcal{P}^{( 1_1)}=\mathcal{P}^{( 1_2)}=1/4$ for the  cavity
occupation probabilities can be  explained considering that at $
t=\pi/(2\Omega_{\rm eff}) $ the system is in the state $
\ket{\psi_2} $ given in \eqref{E3}. Here, the effects of losses
have not been taken into account. (b) Temporal evolution of the
same occupation probabilities $P^{(k)}(t)  $ considered in (a),
after the arrival of a broad Gaussian pulse  exciting the first
cavity when the system is initially prepared in its ground state.
The amplitude and the central frequency of the pulse are $
A/\omega_c = 0.27 $  and
$\omega_d\simeq(\bar{\omega}_{34}+\omega_{50})/2 $, with
$\bar{\omega}_{34}=(\omega_{30} + \omega_{40})/2 $.}} \label{fig4}
\end{figure}

The simultaneous excitation process of the two qubits with a
single photon  can  also be achieved by initially exciting only
the first cavity. This can be realized experimentally by feeding
the cavity with a fast Gaussian pulse (with respect to the
transfer time between the cavities). Specifically, in order to
completely populate the first cavity before the excitation is
transferred to the second one, the pulse bandwidth $ \Gamma =
\pi/\tau$ has to be much larger than the energy splitting $
\Omega=2J $ between the two cavity normal modes, i.e., $ \Gamma
\gg \Omega $. Figure~\ref{fig4}(a) shows the time evolution of the
occupation probabilities $ \mathcal{P}^{( 1_1)} (t)$, $
\mathcal{P}^{( 1_2)} (t)$, and $ \mathcal{P}^{( ee)} (t)$ with the
system initially prepared in the state \be \label{E1}
\ket{\psi_1}=\left(\ket{1_A,0_{S},g,g}+\ket{0_A,1_{S},g,g}
\right)/\sqrt{2}=\ket{1_1}\,. \ee Besides the expected photon
hopping between the two weakly-coupled cavities, as the time
evolves, we observe that this excitation-transfer process is
accompanied by the simultaneous excitation of the two qubits.
However, in contrast to the previous case, where the probability
for the qubits to be simultaneously excited reached one, here we
observe the maximum probability $ \mathcal{P}^{( ee)}=1/2$ at $
t=\pi/(2\Omega_{\rm eff})$. This difference can be explained by
considering that a direct excitation of the first cavity only
corresponds to an equal weight superposition of both the symmetric
and antisymmetric  normal modes (see \eqref{E1}), each of them
carrying half of the total initial excitation.  However, since the
qubits do not interact with the symmetric  mode, only the
antisymmetric excitation contribution can be transferred to the
two qubits, resulting into a maximum joint qubit excitation
probability $ \mathcal{P}^{( ee)}=1/2$.
This result indicates that the simultaneous excitation of two
qubits with one photon is still possible, but the qubits cannot be
excited with probability $ \mathcal{P}^{( ee)}=1$. Moreover, the
corresponding values $ \mathcal{P}^{( 1_1)}=\mathcal{P}^{(
1_2)}=1/4$ for the  cavity occupation probabilities can be
explained by directly following the  dynamics of the system.
Indeed we observe that, when the system is prepared in the
superposition state $\ket{\psi_1}$, then the state
$\ket{0_A,1_{S},g,g} $ evolves freely, as an eigenstate of $
\hat{\mathcal{H}}$. On the contrary, since the qubits are coupled
with the antisymmetric normal mode, once again we observe the
coherent energy exchange process
$\ket{1_A,0_{S},g,g}\leftrightarrow \ket{0_A,0_{S},e,e}$ so that
at $ t=\pi/(2\Omega_{\rm eff}) $ the system is in the state \be
\label{E2}
\ket{\psi_2}=\left(\ket{0_A,0_{S},e,e}+\ket{0_A,1_{S},g,g}
\right)/\sqrt{2}\,. \ee In terms of the energy eigenstates of the
Hamiltonian of the uncoupled system ($ J=g=0 $),  the state $
\ket{\psi_2}  $ can be expressed as \be \label{E3} \ket{\psi_2}=
\frac{1}{\sqrt{2}} \ket{0_1,0_2,e,e} + \frac{1}{2}
\left(\ket{1_1,0_2,g,g}+\ket{0_1,1_2,g,g} \right) \,, \ee thus
explaining the observed values, in Fig.~4(a), for the occupation
probabilities $ \mathcal{P}^{( 1_1)},\mathcal{P}^{( 1_2)}$ and $
\mathcal{P}^{( ee)} $.

Finally, even for this case we consider the system initially
prepared in its ground state and study the system dynamics after
the arrival of a Gaussian pulse exciting the first cavity  and
described by \eqref{pulse1}. Unlike the previous case, in order to
excite the states $ \ket{E_3} $, $ \ket{E_4} $ and $ \ket{E_5} $
we need to apply a broad-bandwidth Gaussian pulse with central
frequency $\omega_d\simeq(\bar{\omega}_{34} + \omega_{50})/2 $,
where $\bar{\omega}_{34}=(\omega_{30} + \omega_{40})/2 $. The
temporal evolution of the occupation probabilities, after the
arrival of the Gaussian pulse feeding the first cavity, is shown
in \figref{fig4}(b). It is interesting to observe that, in the
absence of the qubits, the excitation would  simply be transferred
from one cavity to the another. Indeed, the simultaneous
excitation process can take place  because, except for some
specific instants of time, in which the excitation is totally
localized  in one cavity only, the field is  delocalized over the
two adjacent cavities. For this reason, both qubits feel the
electric field simultaneously  reaching the maximum excitation
probability when the field is equally distributed between the two
cavities, even if they are detuned from the cavity mode.

The causal mechanisms underlying the possibility for a single
photon to be jointly absorbed by the two qubits under different
excitation processes can be explained by considering a simpler
system constituted   by  the two weakly coupled cavities only. If
the system is initially prepared in the state $ \ket{1_1,0_2}$,
the complete population transfer $ \ket{1_1,0_2}\to \ket{0_1,1_2}
$  will take place after a period $ T=\pi/ (2J)  $. From an
experimental point of view, if we consider the system to be
prepared in its ground state $  \ket{0_1,0_2} $, the direct
excitation of the antisymmetric normal mode
$\ket{-}=\left(\ket{1_1,0_2}-\ket{0_1,1_2} \right)/\sqrt{2} $ can
be realized by feeding the first cavity with a Gaussian pulse
whose bandwidth $ \Gamma  $ has to be smaller than the energy
splitting  between the two normal modes, \textit{i.e.},  $ \Gamma
\ll 2J $. Since in this case for the temporal pulse width $ \delta
t \gg T $, the photon can travel back and forth between the two
cavities  before the normal mode becomes fully populated and the
electric field always results  delocalized over the whole cavity
array.

In contrast to this, the realization of a localized cavity mode
(e.g., the state $ \ket{1_1,0_2}  $)  can be achieved by feeding
the first cavity with  a fast optical Gaussian pulse whose
bandwidth has to be large enough to excite the superposition state
$ \left( \ket{+}+\ket{-}\right) /\sqrt{2}  $. The condition $
\Gamma \gg 2J $ ensures that, being $ \delta t \ll T $, the pulse
duration is short enough to localise the electric field in the
first cavity. Then, due to the cavity-cavity interaction the
excitation-transfer process will take place and  the electric
field will be completely delocalized over the two cavities only at
the instants of time when  $\mathcal{P}^{( 1_1)}=\mathcal{P}^{(
1_2)}=1/2$.

\subsection{Three cavities, two qubits and one photon} \label{sec2B}

\begin{figure}[!h]
\centering
\includegraphics[scale=0.48]{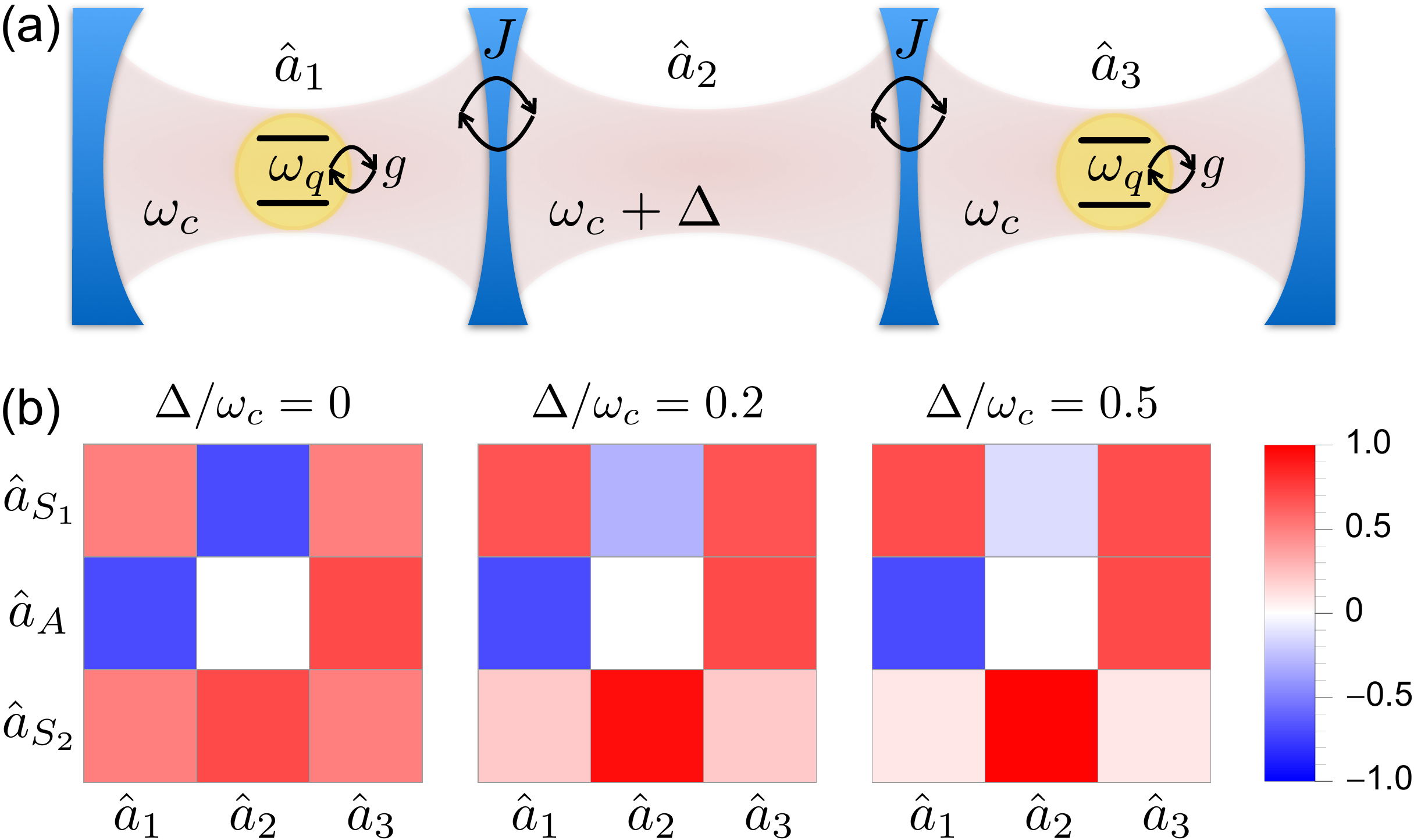}
\caption{(a) Sketch of an array of three optical resonators
weakly coupled with their nearest neighbours. The two resonant end
cavities  ultrastrongly interact with a single qubit with
transition frequency $ \omega_{q} $, while the central cavity is
empty and can be detuned by an amount $\Delta$. The photon hopping
rate between the three resonators and the light-matter coupling
strength are indicated with $ J $ and $ g $, respectively. (b)
Density plot of the transformation matrix diagonalizing
$\hat{H}_{C}^{\prime} $, evaluated for different values of the
normalized detuning $\Delta/\omega_{c} $ for $ J/\omega_{c}=0.05
$ and $\eta\equiv |g|/\omega_{c}=0.3 $. While the antisymmetric
normal mode is not affected by the detuning, the spatial profile
of the two symmetric normal modes changes significantly with
increasing values of $ \Delta $. Specifically, for higher values
of the detuning, the lower-energy mode $ \hat{a}_{\text{S}_1} $
becomes totally delocalized in the end cavities, while the
higher-energy mode $ \hat{a}_{\text{S}_2} $  is completely
localized in the central cavity.} \label{fig5}
\end{figure}

Here we extend the previous analysis to a more complex system
consisting of an array of three  weakly-coupled cavities, where
the two end cavities  ultrastrongly interact with a single qubit
while the central cavity is empty [as shown in \figref{fig5}(a)].
The Hamiltonian describing the system is \bea
\label{h3cav}\nonumber
\hat{H} =\sum_{n=1}^3  \omega_{c}^{(n)} \,\ac_n \an_n +  \sum_{n=1}^2&\biggl[&\omega_{q}\, \sp^{(n)} \sm^{(n)} + J  \left(  \ac_{n} \an_{n+1} + \ac_{n+1} \an_{n} \right)\\
&+&\left| g\right|e^{i \varphi_{(2n-1)}} \hat{X}_{(2n-1)} \left(
\cos\theta\, \sx^{(n)} + \sin \theta\,\sz^{(n)}  \right)\biggr]\,,
\eea where the normalized hopping rate and light-matter coupling
strength are, respectively, $ J/\omega_{\rm c}=0.05 $ and  $
\eta\equiv |g|/\omega_{\rm c}=0.3 $, and the presence of the
longitudinal interaction term is taken into account by considering
a mixing angle $ \theta=\pi/6 $. Due to the small value of the
normalized hopping rate $ J/\omega_{c} $, we apply the RWA to the
cavity-cavity interaction term.  Moreover, we consider the case in
which the two end cavities are resonant
$(\omega_{c}^{(1)}=\omega_{c}^{(3)}=\omega_{c})$, while the
central cavity can be detuned by an amount $\Delta$. The
interaction between these three cavities is described by the
Hamiltonian \be \hat{H}_C^{\prime}=  \omega_{c} \,(\ac_1 \an_1 +
\ac_3 \an_3)  +(\omega_{c}+\Delta ) \,\ac_2 \an_2+ J
\sum\limits_{n=1}^2 (  \ac_{n} \an_{n+1} + \ac_{n+1} \an_{n}) \,,
\ee which produces three normal modes: an antisymmetric flat
state, and two symmetric  modes  whose energy splitting is
$\Omega=\sqrt{8J^2+\Delta^2}  $.
The transformation, which diagonalizes $ \hat{H}_{C}^{\prime} $,
can be written in matrix form: \be \label{bogol}
\begin{pmatrix}
    \hat{a}_{\text{S}_1}\\
    \hat{a}_{A}\\
    \hat{a}_{\text{S}_2}
\end{pmatrix}=\begin{pmatrix}
    2J/\mathcal{N}_{-}  & \bigl(\Delta-\Omega\bigr) / \mathcal{N}_{-} &2J/\mathcal{N}_{-} \\
    -1/\sqrt{2} & 0 & 1/\sqrt{2} \\
    2J/\mathcal{N}_{+}  & \bigl(\Delta+\Omega\bigr) / \mathcal{N}_{+}  & 2J/\mathcal{N}_{+}
\end{pmatrix}
\begin{pmatrix}
    \hat{a}_{1}\\
    \hat{a}_{2}\\
    \hat{a}_{3}
\end{pmatrix}\, ,
\ee with $\mathcal{N}_{\pm} = \left[8J^2+(\Delta\pm\Omega)^2
\right]^{1/2}  $. In \eqref{bogol},  $ \hat{a}_{A}  $ and $
\hat{a}_{\text{S}_1(\text{S}_2)} $ are, respectively, the bosonic
annihilation operators for the antisymmetric state and the two
symmetric normal modes whose corresponding frequencies are
$\omega_A = \omega_c$ and $ \omega_{\text{S}_{1,2}}=\left(2
\omega_{\rm c}+\Delta\mp\Omega\right) /2$.

The effect of the detuning on the spatial profile of the three
normal modes is displayed in \figref{fig5}(b). We observe that,
unlike the antisymmetric mode, which is not affected by the
detuning, the {\em spatial} profile of the two symmetric normal
modes changes significantly with increasing values of $ \Delta $,
and also exhibits an opposite behavior. Specifically, while the
lower-energy mode $ \hat{a}_{\text{S}_1} $ becomes totally
delocalized in the end cavities, the higher-energy mode $
\hat{a}_{\text{S}_2} $  completely localizes in the central
cavity. This preliminary analysis suggests that the choice of
coupling the two qubits to the antisymmetric mode, which is both
detuning-independent and  delocalized in the two end cavities
where the qubits are placed, is  the best strategy for  achieving
the desired effect of  simultaneous excitation of  two qubits with
a single photon. Focusing on the resonant case $ \Delta=0 $, the
Hamiltonian of \eqref{h3cav} can be conveniently rewritten in
terms of the normal mode operators as \bea
\label{h3cavnew}\nonumber
\hat{\mathcal{H}}^{\prime}&=&\sum_{k\in[\text{S}_1,\text{S}_2,A]}^{}
\omega_{k} \,\hat{a}_{k}^{\dag}  \hat{a}_{k}^{}+\,
 \omega_{q}\sum_{n=1}^2  \sp^{(n)} \sm^{(n)}\\
 &+& |g|\left[\frac{1}{\sqrt{2}}\left( \hat{X}_{\text{S}_2} - \hat{X}_{\text{S}_1}\right)  \left(\cos\theta \, \hat{\Phi }_{x}^{+} +  \sin \theta\, \hat{\Phi }_{z}^{+}  \right) -  \hat{X}_{A} \left(\cos\theta \, \hat{\Phi }_{x}^{-} +  \sin \theta\, \hat{\Phi }_{z}^{-} \right)\right] \,,
\eea where $ \hat{X}_{A}\equiv \hat{a}^\dag_{A} + \hat{a}_{A}^{}$,
$ \hat{X}_{\text{S}_1(\text{S}_2)}\equiv
\hat{a}^\dag_{\text{S}_1(\text{S}_2)} +
\hat{a}_{\text{S}_1(\text{S}_2)}^{}$, and $\hat{\Phi
}_{x(z)}^{\pm}\equiv \left( e^{i
\varphi_1}\,\hat{\sigma}_{x(z)}^{(1)} \pm e^{i \varphi_3}\,
\hat{\sigma}_{x(z)}^{(2)}\right)/\sqrt{2}$.

\begin{figure}[!h]
\centering
\includegraphics[scale=0.42]{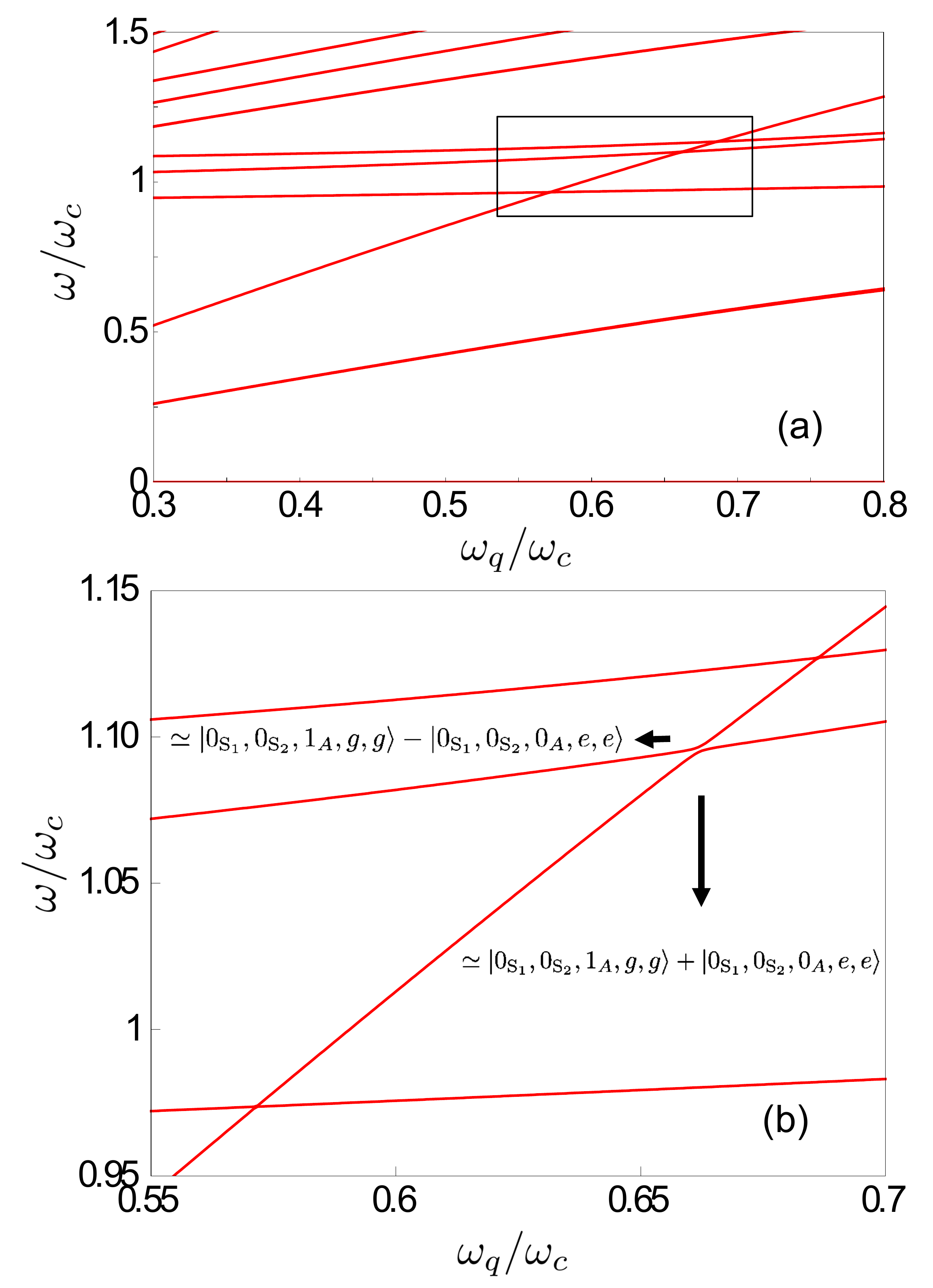}
\caption{\small{Energy differences $
\omega_{i0}=\omega_{i}-\omega_{0}$ for the lowest-energy dressed
states of $ \mathcal{H}^{\prime} $ as a function of the normalized
qubit frequency $ \omega_{q}/\omega_{c}$ in the absence of
detuning ($ \Delta=0 $). We set the normalized qubit-resonator
coupling rate and the inter-cavity photon hopping rate to  $ \eta
\equiv|g|/\omega_{c}=0.3 $ and $ J/\omega_{c}=0.05 $,
respectively. The phases for the cavity-qubit coupling strengths
are  $\varphi_{1}=0  $ and $\varphi_{3}=\pi$, while the
longitudinal interaction coupling term is included by considering
a mixing angle $ \theta= \pi/6 $. (b) Enlarged view of the inset
in (a). When the cavity-qubit coupling strengths are opposite in
phase ($\varphi_{1}=0,\varphi_{3}=\pi $), the avoided level
crossing  results from the coupling between the states
$\ket{0_{\text{S}_1},0_{\text{S}_2},1_{A},g,g}$ and
$\ket{0_{\text{S}_1},0_{\text{S}_2},0_{A},e,e}$,  due to the
presence of counter-rotating terms in the system Hamiltonian. The
energy splitting reaches its minimum at $\omega_{q}  \simeq
\omega_{A}/2 $. The presence of two energy-level crossings
corresponding to the qubit frequencies $ \omega_{q} \simeq
\omega_{\text{S}_1}/2  $ and   $ \omega_{q} \simeq
\omega_{\text{S}_2}/2  $, both involving the state
$\ket{0_{\text{S}_1},0_{\text{S}_2},0_{A},e,e}$,  indicates that
the two  qubits do not couple with the two symmetric normal modes.
}} \label{fig6}
\end{figure}

Figure \ref{fig6}(a) shows the energy differences $
\omega_{i0}=\omega_{i}-\omega_{0}$ for the lowest-energy
eigenstates $ \ket{E_i} $ (with $ i = 0,1, \dots $) of
$\hat{\mathcal{H}}^{\prime} $, numerically calculated  as a
function of the normalized qubit frequency $
\omega_{q}/\omega_{c}$ by setting the  phases  of the two
cavity-qubit coupling strengths to $  \varphi_{1}=0 $ and
$\varphi_{3}=\pi $, respectively. Here we use the notation
$\ket{\mathcal{N}_{\text{S}_1},\mathcal{N}_{\text{S}_2},\mathcal{N}_A,q_1,q_2}=\ket{\mathcal{N}_{\text{S}_1}}\bigotimes\ket{\mathcal{N}_{\text{S}_2}}\bigotimes\ket{\mathcal{N}_A}
\bigotimes\ket{q_1}\bigotimes \ket{q_2} $ for the eigenstates $
\ket{E_i} $, where $q=\left \{g,e\right \}$ denote the qubit
ground or excited states, respectively, and
$\ket{\mathcal{N}_{k}}=\left \{\ket{0},\ket{1},\ket{2},\dots
\right \} $, with $ k\in[\text{S}_1,\text{S}_2,A] $, represents
the Fock state with photon occupation $ \mathcal{N}_{k}$ in the
corresponding normal mode.

Note that the energy spectrum presents a more complicated
structure with respect to the case of the two cavity-qubit array.
Specifically, two energy-level crossings, both involving the state
$\ket{0_{\text{S}_1},0_{\text{S}_2},0_{A},e,e}$ (which displays a
linear behavior with $\omega \approx  2 \omega_{q}$), can be
observed in correspondence to the qubit frequencies $ \omega_{q}
\simeq \omega_{\text{S}_1}/2  $ and   $ \omega_{q} \simeq
\omega_{\text{S}_2}/2  $. The other states involved in the two
energy-level crossings are, respectively,
$\ket{1_{\text{S}_1},0_{\text{S}_2},0_{A},g,g}$  and
$\ket{0_{\text{S}_1},1_{\text{S}_2},0_{A},g,g}$, indicating that
when the cavity-qubit coupling strengths have opposite phases, the
qubits do not couple with the two symmetric normal modes.
Interestingly, in the region between these two-level crossings, an
apparent additional  one between the levels $ \ket{E_{4}} $ and $
\ket{E_{5}} $  appears at $ \omega_{q} \simeq \omega_{A}/2$.
Actually, what appears as a crossing on this scale turns out to be
an avoided-level crossing on an enlarged view, as in
\figref{fig6}(b).
This splitting, which  has a normalized value $ 2 \Omega_{\rm eff
}/\omega_{\rm c}=2\times 10^{-3} $ at its minimum, clearly
originates from the hybridization of the states
$\ket{0_{\text{S}_1},0_{\text{S}_2},0_{A},e,e}$  and
$\ket{0_{\text{S}_1},0_{\text{S}_2},1_{A},g,g}$.  The resulting
states are well approximated by \be
\ket{E_{4(5)}}=(\ket{0_{\text{S}_1},0_{\text{S}_2},1_{A},g,g}\pm\ket{0_{\text{S}_1},0_{\text{S}_2},0_{A},e,e})/\sqrt{2}\,.
\ee It is important to observe that, similarly to the two
cavity-qubit array case, the coherent coupling between these two
states  would neither be allowed within the RWA, nor in the
absence of the longitudinal interaction term $ (\theta=0) $.
Moreover, the states
$\ket{0_{\text{S}_1},0_{\text{S}_2},0_{A},e,e}$  and
$\ket{0_{\text{S}_1},0_{\text{S}_2},1_{A},g,g}$  do not couple
directly, but the process  occurs via intermediate  energy
non-conserving processes enabled by the counter-rotating terms in
$ \hat{\mathcal{H}}^{\prime} $. It is interesting to observe that,
when  the  coupling strengths are opposite in phase,  all the
possible intermediate virtual transitions for the process
$\ket{0_{\text{S}_1},0_{\text{S}_2},1_{A},g,g}\to
\ket{0_{\text{S}_1},0_{\text{S}_2},0_{A},e,e} $ are induced by the
interaction term proportional to $ \hat{X}_{A} $, while the term
proportional to $\left( \hat{X}_{\text{S}_2} -
\hat{X}_{\text{S}_1}\right) $ gives vanishing contributions. The
choice of same coupling strength  phases  $ \left( \varphi_1 =
\varphi_3\right) $ would lead to the complementary situation with
the two qubits  decoupled from the antisymmetric mode  and
simultaneously interacting with the two  symmetric modes  $
\hat{a}_{\text{S}_1} $ and $ \hat{a}_{\text{S}_2} $. In this case,
the simultaneous excitation of the two qubits with a single photon
can occur via two different processes ($
\ket{1_{\text{S}_1},0_{\text{S}_2},0_{A},g,g}\to
\ket{0_{\text{S}_1},0_{\text{S}_2},0_{A},e,e} $  and
$\ket{0_{\text{S}_1},1_{\text{S}_2},0_{A},g,g}\to
\ket{0_{\text{S}_1},0_{\text{S}_2},0_{A},e,e}$). However, since
the symmetric normal modes are mainly localized in the central
empty cavity, the energy splittings of the  avoided-level
crossings between these states, as well as the corresponding
effective couplings, are much smaller.

\subsubsection{Dynamics after exciting the antisymmetric mode}

In order to fully understand the excitation transfer between a
single photon and two qubits in a three cavity-qubit array, we now
study the dynamics of the system  initially prepared in the
one-photon state
$\ket{1_A}\equiv\ket{0_{\text{S}_1},0_{\text{S}_2},1_{A},g,g}$,
fixing  the qubit frequency  at the value where the splitting in
\figref{fig6}(b) between levels $ \ket{E_{4}} $ and $ \ket{E_{5}}
$ is minimum. Moreover, for the sake of simplicity, we consider
the system loss rates to be significantly smaller than the
frequency splitting between the levels involved in the
avoided-level crossing, so that the effect of dissipation can be
neglected.

\begin{figure}[h]
    \centering
    \includegraphics[scale=0.21]{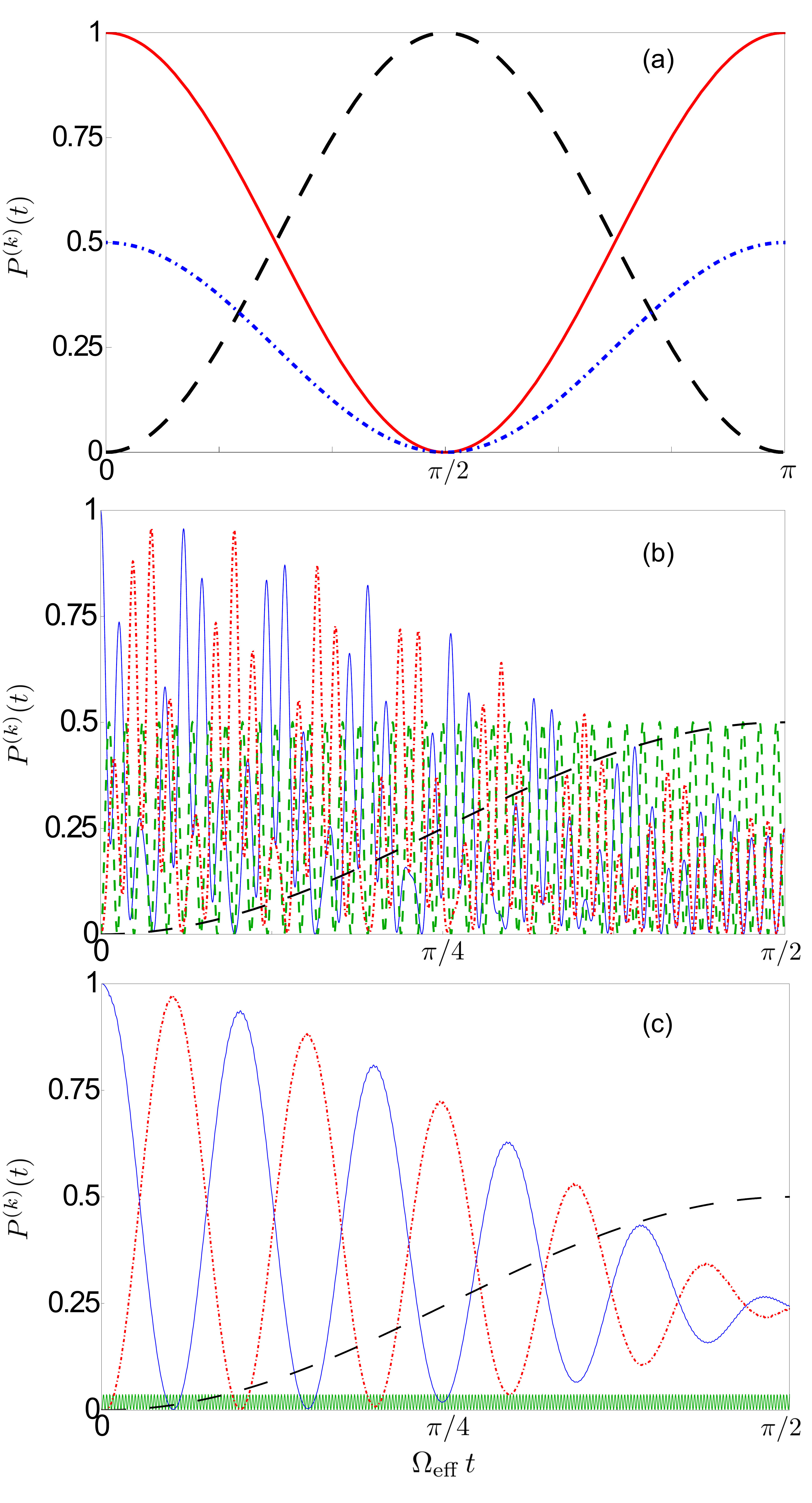}
    \caption{(a) Time evolution of the occupation probabilities $P^{(k)}(t)\equiv \langle \hat{\mathcal{P}}_{k} \rangle $, with $\hat{\mathcal{P}}_{k}= \ketbra{k}{k}  $, for the  one-photon states $\ket{1_A}$ (red solid curve) and $ \ket{1_1} $ (blue dot-dashed curve), together with the probability  $ \mathcal{P}^{(ee)} (t)$ of having both qubits simultaneously excited (black dashed curve). Here, the system is initially prepared in the one-photon state $\ket{1_A}\equiv\ket{0_{\text{S}_1},0_{\text{S}_2},1_{A},g,g}$ and in the absence of detuning ($ \Delta=0 $).
        As the times evolves, the excitation is progressively transferred to the two qubits at the same time, until the maximum simultaneous qubit excitation ($ \mathcal{P}^{( ee)}=1  $) by a single photon is reached at $ t=\pi/(2 \Omega_{\rm eff}) $.
        (b) Temporal dynamics of the occupation probabilities $P^{(1_1)}$ (blue solid curve), $P^{(1_2)}$ (green dashed curve) and $P^{(1_3)}$ (red dot-dashed curve), together with the probability  $ \mathcal{P}^{(ee)} $ (black dashed curve)  of having both qubits simultaneously excited when the system is initially prepared in the one-photon state $\ket{\psi_1} =\ket{1_1}$ of the first resonator with $\Delta=0$.
        (c)  Temporal dynamics of the same occupation probabilities $P^{(k)}(t) $ considered in (b) but in the presence of the detuning $\Delta/\omega_{c}=0.5$.
    }\label{fig7}
\end{figure}
\clearpage 

The numerically calculated time evolution of the occupation probabilities $P^{(k)}(t)\equiv \langle \hat{\mathcal{P}}_{k} \rangle $, with $\hat{\mathcal{P}}_{k}= \ketbra{k}{k}  $, for the  one-photon states $\ket{1_A}$ (a single photon in the antisymmetric normal mode), $ \ket{1_1} $ and $ \ket{1_3} $  (a single photon in the first and third cavities, respectively), together with the probability  $ \mathcal{P}^{(ee)} (t)$ of having both qubits simultaneously excited, are displayed in \figref{fig7}(a). 
As expected, $\mathcal{P}^{( 1_{ A})} (0)=1$ at the initial
instant of time, and  since
$\ket{1_A}=\left(\ket{0_1,0_2,1_3}-\ket{1_1,0_2,0_3}
\right)/\sqrt{2} $, we observe that the end cavities are equally
populated $\left[\mathcal{P}^{( 1_1)} (0)=\mathcal{P}^{( 1_3)}
(0)=1/2\right]$, while the central cavity is empty $ \left[
\mathcal{P}^{( 1_2)} (0)=0\right]  $. As time evolves, the
excitation is progressively transferred to the two qubits at the
same time, until the maximum simultaneous qubit excitation [with
$\mathcal{P}^{( ee)} (0)=1$] is reached at $ t=\pi/(2 \Omega_{\rm
eff}) $. It is interesting to observe that, since we are exciting
the antisymmetric mode, the central cavity remains empty during
the whole process. This fact remains valid even if the central
cavity is detuned $ \left( \Delta\ne 0\right)$, so that  the
system displays the same dynamics in the non-zero detuning case,
the only difference being that the simultaneous excitation of the
two qubits is achieved at a different instant of time.
The excitation of the antisymmetric normal mode, which is the most
effective way to achieve the desired effect, can be experimentally
achieved by sending  a suitable narrow  Gaussian pulse to the
first cavity, whose central frequency of the pulse has to be
chosen to be in the middle of the two split transition energies
$\omega_d=(\omega_{40}+\omega_{50})/2 $. For the sake of
simplicity, here we do not present numerical calculations for the
dynamics of the system excited by a Gaussian pulse, since the
results would not add any additional physical information.
\subsubsection{Dynamics after exciting the first cavity}

We now turn to the study of the system dynamics when, instead of
exciting the antisymmetric mode,  we  directly excite only the
first cavity. Unlike the previous case, this  process strongly
depends on the detuning $ \Delta $.

The dynamics of the occupation probabilities  $\mathcal{P}^{(k)} $
for a three coupled cavity-qubit array system initially prepared
in the one-photon state $ \ket{1_1} $ can be studied for arbitrary
values of the detuning $ \Delta $ by using a semi analytical
approach. Choosing  the energy eigenstates $ \ket{E_j} $ of $
\hat{\mathcal{H}}^{\prime}$ as basis, and considering that only
the states $ \ket{E_3},\ket{E_4},\ket{E_5} $, and $ \ket{E_6} $
are involved in the process, the time evolution of the initial
state $ \ket{1_1(t)} $ in the Schr\"odinger picture  can be
written as: \be
 \ket{1_1(t)}=\sum\limits_{j=3}^{6} c_{j}\,e^{-i \omega_{j}t}\ket{E_j},
\ee where $ \omega_{j} $ are the numerically evaluated eigenvalues
and the coefficients $c_{j} $, which depend on $ J$ and $\Delta $,
are given by the elements of the transformation matrix in
\eqref{bogol}. The time evolution of the occupation probability
$\mathcal{P}^{(k)}(t) $ for a generic state $
\ket{k}=\sum\limits_{j=3}^{6} d_{j}\ket{E_j}  $ are simply given
by: \be \mathcal{P}^{(k)}(t)=\left | \braket{k}{1_1(t)} \right |^2
= \left |\sum\limits_{j=3}^{6} c_{j} d_{j} \,e^{-i \omega_{j}t}
\right |^2\,. \ee Figure~\ref{fig7}(b) shows the time evolution of
the occupation probabilities  $ \mathcal{P}^{( 1_1)} (t)$, $
\mathcal{P}^{( 1_2)} (t)$, $ \mathcal{P}^{( 1_3)} (t)$, and $
\mathcal{P}^{( ee)} (t)$ for the resonant case $ \Delta=0 $, when
the system is initially prepared in the one-photon state \be
\label{E4}
\ket{\psi_1}=\left(\ket{0_{\text{S}_1},1_{\text{S}_2},0_A,g,g}-\ket{1_{\text{S}_1},0_{\text{S}_2},0_A,g,g}
\right)/2 -\ket{0_{\text{S}_1},0_{\text{S}_2},1_A,g,g}/\sqrt{2}
=\ket{1_1}\,. \ee As the time evolves, we observe that the
excitation, continuously propagating between the three cavities,
is progressively transferred to both qubits at the same time, even
if they are detuned from the central cavity. We observe that
initially the system dynamics is not affected by the presence of
the qubits and  the system behaves like an array of three
weakly-coupled resonant cavities. Once the excitation starts to be
transferred to the qubits, the system undergoes a more complex
dynamics and the probability for the qubits to be simultaneously
excited reaches its maximum  $ \mathcal{P}^{( ee)}=1/2$ at $
t=\pi/ (2 \Omega_{\rm eff }) $. As expected, at this instant of
time the electric field is completely delocalized  only in the end
cavities with $\mathcal{P}^{(1_1)}=\mathcal{P}^{( 1_3)}=1/4$,
while the central cavity is depopulated. This result can be easily
understood by considering that, after the system is prepared in
the state $\ket{\psi_1}  $, the superposition $
\left(\ket{0_{\text{S}_1},1_{\text{S}_2},0_A,g,g}-\ket{1_{\text{S}_1},0_{\text{S}_2},0_A,g,g}
\right)/2  $ evolves freely since the qubits are not coupled to
the symmetric modes.

In contrast to this, due to the interaction between the qubits and
the antisymmetric mode, the above-described
coherent-energy-exchange process $
\ket{0_{\text{S}_1},0_{\text{S}_2},1_A,g,g} \to
\ket{0_{\text{S}_1},0_{\text{S}_2},0_A,e,e} $ takes place so that
at $ t=\pi/ (2 \Omega_{\rm eff }) $  the system will be in the
state \be \label{E5}
\ket{\psi_2}=\left(\ket{0_{\text{S}_1},1_{\text{S}_2},0_A,g,g}-\ket{1_{\text{S}_1},0_{\text{S}_2},0_A,g,g}
\right)/2 -\ket{0_{\text{S}_1},0_{\text{S}_2},0_A,e,e}/\sqrt{2}\,.
\ee The observed values for the occupation probabilities
$\mathcal{P}^{(k)} $ in Fig.~7(b) can be explained by considering
that, in terms of the  energy eigenstates of the Hamiltonian of
the uncoupled system $ (J=g=0) $, this state can be expressed as
\be \label{E6}
\ket{\psi_2}=\left(\ket{1_{1},0_{2},0_3,g,g}+\ket{0_{1},0_{2},1_3,g,g}
\right)/2 -\ket{0_1,0_2,0_3,e,e}/\sqrt{2}\,. \ee

Finally, in \figref{fig7}(c) we present numerical results for the
dynamics of the system initially prepared in the state $
\ket{\psi_1}  $ for $ \Delta/\omega_c=0.5 $. In this case, the
excitation of the first cavity can be obtained by exciting a
superposition of the lowest-energy symmetric mode and the
antisymmetric mode [see \figref{fig5}(b)]. This could be
experimentally realized by sending a suitable broad Gaussian pulse
to the first cavity, able to excite the energy levels $
\ket{E_{3}} $, $ \ket{E_{4}} $, $ \ket{E_{5}} $ and $ \ket{E_{6}}
$ simultaneously.

We observe that the system dynamics displays a different trend
with respect to the resonant case. Indeed, due to the strong
detuning, the central cavity acts like a high-potential barrier
and the excitation is transferred back and forth via photon
tunneling only between the end cavities,  with the system
effectively behaving like a two coupled cavity-qubit array. During
the whole process, the central cavity remains very low-populated
and the excitation is progressively transferred to the qubits, and
the maximum simultaneous excitation probability  $ \mathcal{P}^{(
ee)}=1/2$ is reached at $ t=\pi/ (2 \Omega_{\rm eff }) $, where $
\Omega_{\rm eff }/\omega_{c}= 4.5 \times 10^{-4} $ is the
effective coupling for $ \Delta/\omega_c=0.5 $ between two qubits
and a single photon.

The processes described here could be experimentally observed by
placing two superconducting artificial atoms at opposite ends of
an array of capacitively-coupled superconducting waveguides. These
anomalous multiatom excitation and emission processes can find
applications for the development of novel quantum technologies for
quantum information and communication as, for example, the
realization of new effective methods for quantum information
transfer between photons and qubits in quantum networks.

\section{Conclusions} \label{sec3}

When the light-matter coupling strength increases, the vacuum
fluctuations of the electromagnetic field become able to
efficiently induce virtual transitions, replacing the role of the
intense applied fields in nonlinear
optics~\cite{Kockum2017a,Forn-Diaz2018,Kockum2019}. In this way,
higher-order processes involving counter-rotating terms can create
an effective coupling between two states of a system with
different numbers of excitations. One of the most interesting
examples is the process where a single photon in an
electromagnetic resonator can jointly excite two atoms interacting
with the same resonator. We have investigated this intriguing
nonlinear optical process in the case where each of the two atoms
is coupled to a distinct resonator. Specifically we studied: (i)
the case of two coupled resonators, each of them interacting with
a single atom, and (ii) the case of two resonator-atom systems
weakly coupled through a central resonator (resonant or detuned
with respect to the other two). We studied the dynamics of these
systems under different excitation conditions, showing  that a
coherent energy transfer between a single photon and two spatially
separated atoms can still occur with a probability approaching one,
under specific excitation conditions.

The  results obtained provide interesting insights into subtle
causality issues underlying the simultaneous excitation of
two-level systems placed in { spatially separated} resonators. The processes
described here could be experimentally realized in
state-of-the-art circuit QED systems.
{Specifically, in our calculations we used a normalized coupling strength $\eta =|g|/\omega_c = 0.3$. It is not a problem to experimentally reach such a value in circuit QED systems.
Using a flux qubit coupled to an $LC$ oscillator, values of $\eta$ of the order of $\simeq 0.65$, and even beyond it, have been obtained. In order to observe the dynamics shown, e.g., in \figref{fig3}, the broadenings should be significantly lower than the  splitting shown in \figref{fig2}(b). Such a splitting is of the order $\Delta / \omega_c \sim  10^{-3}$. For a typical resonance frequency $\omega_c / 2 \pi \simeq 6$ GHz, the splitting is thus of the order of $0.35$ GHz. In circuit QED systems, typical damping rates are orders of magnitude lower than this value \cite{Gu2017}. Notice also that such a splitting can be increased by increasing $\eta$; it scales as $\eta^3$ \cite{Garziano2016}.}
{Recently, the possibility to coherently couple distant ($\sim 1$ m) cavity-QED systems, connecting them by a {microwave cable} with negligible loss \cite{Leung2019,Kato2019} has been experimentally demonstrated. These  results opens the possibility to implement the process proposed here involving  distant artificial atoms.}

{It would be interesting and
useful for applications to extend these studies to the cases of
spatially separated atomic ensembles~\cite{Macri2020}, and distant
qubits coupled to open waveguides~\cite{Forn-Diaz2017}. Moreover,
further insights on the processes here analysed can be gained by
considering  two or more  spatially separated atoms in a multimode
cavity~\cite{SanchezMunoz2018}. It would also be interesting to
extend the description of  other vacuum-boosted nonlinear optical
processes in the USC regime~\cite{Stassi2017,Kockum2017a},
including coupled and/or multi-mode resonators.}
\newpage
{\Large \bf Acknowledgments}

F.N. is supported in part by: NTT Research, 
Army Research Office (ARO) (Grant No. W911NF-18-1-0358), 
Japan Science and Technology Agency (JST) 
(via the Q-LEAP program, and the CREST Grant No. JPMJCR1676), 
Japan Society for the Promotion of Science (JSPS) (via the KAKENHI Grant No. JP20H00134 
and the JSPS-RFBR Grant No. JPJSBP120194828), 
the Asian Office of Aerospace Research and Development (AOARD), 
and the Foundational Questions Institute Fund (FQXi) via Grant No. FQXi-IAF19-06.
 S.S. was supported by the Army Research Office (ARO)
(Grant No. W911NF1910065). A.M. was supported by the Polish
National Science Centre (NCN) under the Maestro Grant No.
DEC-2019/34/A/ST2/00081.
G.F. was supported by the QuantERA grant SiUCs by CNR GA nr. 
731473 QuantERA,
and by University of Catania, Piano di Incentivi per la Ricerca di Ateneo 2020/2022, proposal Q-ICT.

\vspace{1 cm}


\begin{center}

\Large{\bf References}
\end{center}

\bibliographystyle{naturemag}


\end{document}